\documentclass[prd,a4paper,showpacs]{revtex4}
\usepackage{epsfig}
\usepackage{color}
\usepackage{amssymb}
\usepackage{amsfonts}
\usepackage{graphicx}
\usepackage{keyval,graphicx}
\usepackage{textcomp,wasysym}

\newcommand{\veps}{\mbox{\boldmath $\epsilon$\unboldmath}}

\begin{document}
\title{Understanding the radiative decays of vector charmonia to light pseudoscalar mesons}
\author{Qiang Zhao$^{1,2}$\footnote {E-mail: zhaoq@ihep.ac.cn}}

\affiliation{1) Institute of High Energy Physics, Chinese Academy of
Sciences, Beijing 100049, P.R. China \\
2) Theoretical Physics Center for Science Facilities, CAS, Beijing
100049, P.R. China}

\begin{abstract}

We show that the newly measured branching ratios of vector charmonia
($J/\psi$, $\psi^\prime$ and $\psi(3770))$ into $\gamma P$, where
$P$ stands for light pseudoscalar mesons $\pi^0$, $\eta$, and
$\eta^\prime$, can be well understood in the framework of vector
meson dominance (VMD) in association with the
$\eta_c$-$\eta(\eta^\prime)$ mixings due to the axial gluonic
anomaly. These two mechanisms behave differently in $J/\psi$ and
$\psi^\prime\to \gamma P$. A coherent understanding of the branching
ratio patterns observed in $J/\psi(\psi^\prime)\to \gamma P$ can be
achieved by self-consistently including those transition mechanisms
at hadronic level. The branching ratios for $\psi(3770)\to\gamma P$
are predicted to be rather small.

\end{abstract}

\date{\today}





\maketitle

\section{Introduction}

The recent measurements of the vector charmonium radiative decays to
light pseudoscalars, i.e. $J/\psi$, $\psi^\prime$ and $\psi(3770)\to
\gamma \pi^0$, $\gamma \eta$ and $\gamma\eta^\prime$, have brought
surprises and interests to us. Earlier, the CLEO
Collaboration~\cite{:2009tia} renewed the branching ratios for
$J/\psi\to \gamma \pi^0$, $\gamma\eta$, $\gamma \eta^\prime$, and
$\psi^\prime\to \gamma\eta^\prime$, which are consistent with the
averages from 2008 Particle Data Group~\cite{Amsler:2008zzb}. The
branching ratio upper limits for $\psi^\prime\to \gamma\pi^0$ and
$\gamma\eta$ were set, which were more than one order of magnitude
smaller than that for $\psi^\prime\to \gamma\eta^\prime$. Meanwhile,
the upper limits for $\psi(3770)\to \gamma P$, where $P$ stands for
pseudoscalar $\pi^0$, $\eta$ and $\eta^\prime$, were set to be about
$10^{-5}\sim 10^{-6}$. The $\psi^\prime$ radiative decays are also
investigated by the BESIII Collaboration with the newly collected
106 million $\psi^\prime$ events, and the results turn out to be
tantalizing. It shows that the branching ratios for $\psi^\prime\to
\gamma\pi^0$ and $\gamma\eta$  are only at an order of $10^{-6}$,
which are nearly two orders of magnitude smaller than
$\psi^\prime\to \gamma\eta^\prime$~\cite{bes-iii}.

The mysterious aspects somehow are correlated with the $J/\psi$ and
$\psi^\prime$ data. It is found that the branching ratio for
$J/\psi\to \gamma\pi^0$ is much smaller than those for $J/\psi\to
\gamma\eta$ and
$\gamma\eta^\prime$~\cite{Amsler:2008zzb,Nakamura:2010zzi}. This
could be a consequence of suppressions of gluon couplings to
isovector currents. As a comparison, the observation in
$\psi^\prime\to \gamma P$ is indeed puzzling. The immediate question
is, what drives the difference of decay patterns between $J/\psi$
and $\psi^\prime$.

In the literature, the radiative decays of the vector charmonia
attracted a lot of theoretical efforts. An early study by the QCD
sum rules~\cite{Novikov:1979uy} suggested the dominance of
short-distance $c\bar{c}$ annihilations. The gluon and $q\bar{q}$
transition matrix elements were computed by coupling the gluon
fields to the pseudoscalar states with which the branching ratio
fraction $BR(J/\psi\to \gamma \eta^\prime)/BR(J/\psi\to \gamma\eta)$
was satisfactorily described. In Ref.~\cite{Chao:1990im}, the
$\eta_c$ mixings with the light pseudoscalars $\eta$ and
$\eta^\prime$ were extracted through the axial gluonic anomaly on
the basis of chiral and large $N_c$ approach. By assuming that the
partial widths of $J/\psi\to \gamma\eta$ and $\gamma\eta^\prime$
were saturated by the $\eta_c$-$\eta(\eta^\prime)$ mixing, the
branching ratios for $J/\psi\to \gamma\eta$ and $\gamma\eta^\prime$
were accounted for to the correct orders of magnitude. This issue
was revisited by Feldmann {\it et al.} who proposed to extract the
mixing and decay constants on the quark flavor
basis~\cite{Feldmann:1998sh}. This scheme can be easily extended to
accommodate the mixing of $\eta_c$ with $\eta$ and $\eta^\prime$
from which the $\eta_c$-$\eta(\eta^\prime)$ mixing angles were
extracted and turned out to be consistent with those from
Refs.~\cite{Chao:1990im,ali,petrov}.

Interestingly, the new data from BESIII for $\psi^\prime\to \gamma
P$ seem to suggest a deviation from the saturation assumption. It
implies that some other mechanisms become important in
$\psi^\prime\to \gamma P$, although they may not play a significant
role in $J/\psi\to \gamma P$. In this work, we shall show that the
vector meson dominance (VMD) model is an ideal framework to make a
coherent analysis of the $\eta_c$-$\eta(\eta^\prime)$ mixing effects
and contributions from intermediate vector mesons. We shall show
that the $\psi^\prime\to \gamma P$ is not saturated by the
$\eta_c$-$\eta(\eta^\prime)$ mixing. Instead, one important
mechanism that drives the difference between $J/\psi$ and
$\psi^\prime\to \gamma P$ and produces the observed patterns is the
sizeable coupling of $\psi^\prime\to J/\psi P$.

As follows, we first give a brief introduction to the VMD model and
lay out the correlated aspects of the $\eta_c$-$\eta(\eta^\prime)$
mixings in Sec. II. The detailed analysis, calculation results and
discussions will then be presented in Sec. III. A brief summary will
be given in Sec. IV.

\section{VMD model and $\eta_c$-$\eta(\eta^\prime)$ mixings}

In the VMD model (e.g. see review of
Refs.~\cite{Bauer:1975bw,Bauer:1977iq}) the electromagnetic (EM)
current can be decomposed into a sum of all neutral vector meson
fields including both isospin-0 and isospin-1 components. The
leading $V\gamma^*$ effective coupling can be written as:
\begin{equation}
\mathcal L
_{V\gamma} = \sum_V \frac{e M_V^2}{f_V} V_\mu A^\mu, \label{lag-1}
\end{equation}
where $V^\mu(=\rho,\omega,\phi, J/\psi, ...)$ denotes the vector
meson field.  The photon-vector-meson coupling constant $e
M_V^2/f_V$ can be extracted  from the partial decay width
$\Gamma_{V\to e^+e^-}$. Neglecting the mass of electron and
positron, we have
\begin{equation}
 \frac{e}{f_V} =
\left[\frac{3 \Gamma_{V\to e^+e^-}}{2 \alpha_e |{\bf
p}_e|}\right]^{\frac{1}{2}}, \label{fv}
\end{equation}
where ${\bf p}_e$ is the electron three-vector momentum in the
vector meson rest frame, and $\alpha_e$ is the EM fine-structure
constant.

\begin{figure}[htb]
\includegraphics[width=0.6\hsize]{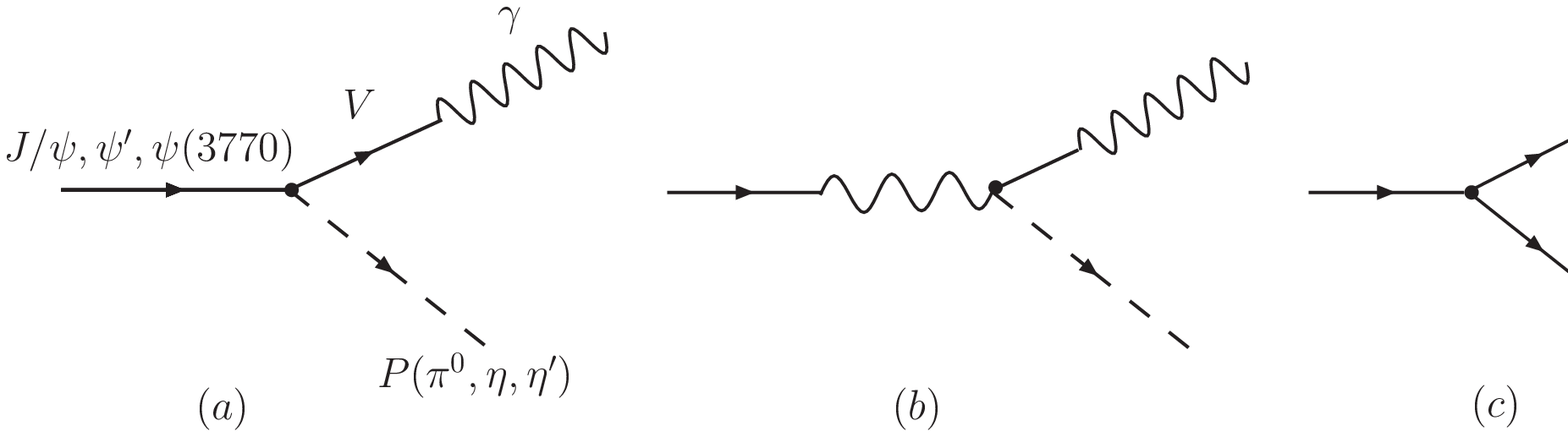}
\caption{ Schematic diagrams for $J/\psi (\psi^\prime, \
\psi(3770))\to \gamma P$ in the frame of VMD. }\label{fig-1}
\end{figure}

For the decays of $J/\psi (\psi^\prime, \ \psi(3770))\to \gamma P$,
the VMD contributing diagrams are illustrated in Fig.~\ref{fig-1}.
This classification is based on the photon producing mechanisms and
related to the experimental measurements. For instance,
Fig.~\ref{fig-1}(a) identifies such a process that the photon is
connected to a hadronic vector meson fields. It requires a sum over
all strong transitions of $J/\psi(\psi^\prime, \ \psi(3770))\to VP$
channels.

The second process in Fig.~\ref{fig-1}(b) is via charmonium
electromagnetic (EM) annihilations. Such a process generally has
small contributions in comparison with the strong transitions.
However, it is likely that the EM amplitudes may have significant
effects in some exclusive decay channels. In recent series
studies~\cite{Li:2007ky,Zhao:2006gw,Zhang:2009kr,Zhao:2010ja} it
shows that in the hadronic decays of $J/\psi (\psi^\prime)\to VP$,
the short (via three gluon annihilation) and long-distance
(Fig.~\ref{fig-1}(c)) transition amplitudes may have a destructive
interfering mode that would efficiently reduce the strong transition
amplitudes in some exclusive channels. As a consequence, the EM
amplitudes may become compatible with the strong ones, and manifest
themselves in experimental observables. This issue is related to the
so-called ``$\rho\pi$ puzzle", which questions why the branching
ratio fraction $BR(\psi^\prime\to \rho\pi)/BR(J/\psi\to \rho\pi)$ is
so strongly suppressed in comparison with the pQCD expectation
values~\cite{Brodsky:1981kj,Chernyak:1981zz,Chernyak:1983ej}. A
review of this subject and some recent progresses on this problem
can be found in the
literature~\cite{Mo:2006cy,Li:2007ky,Zhao:2008eg}.

In the present work, our attention is to understand whether the data
for $J/\psi(\psi^\prime, \ \psi(3770))\to \gamma P$ are consistent
with those for $J/\psi(\psi^\prime, \ \psi(3770))\to VP$, and what
drives the different radiative decay patterns between $J/\psi$ and
$\psi^\prime$. We shall adopt the available experimental
measurements of $J/\psi(\psi^\prime, \ \psi(3770))\to VP$ in the
calculations of the VMD contributions. This means we need not worry
about the detailed transition mechanisms for $J/\psi(\psi^\prime, \
\psi(3770))\to VP$ at this moment. Also, by adopting the
experimental data for $J/\psi(\psi^\prime, \ \psi(3770))\to VP$, we
need not consider the $\eta-\eta^\prime$ mixing processes since they
have been contained in the data for $J/\psi(\psi^\prime, \
\psi(3770))\to VP$.

It is worth noting in advance another feature with this
classification of Fig.~\ref{fig-1}. Namely, transitions between
vector charmonia may also contribute. For instance, $\psi^\prime\to
J/\psi\eta$ will contribute to $\psi^\prime\to \gamma\eta$. We will
show later that this process is essential for understanding the
radiative decay patterns for $J/\psi$ and $\psi^\prime\to \gamma P$.

\begin{figure}[htb]
\includegraphics[width=0.3\hsize]{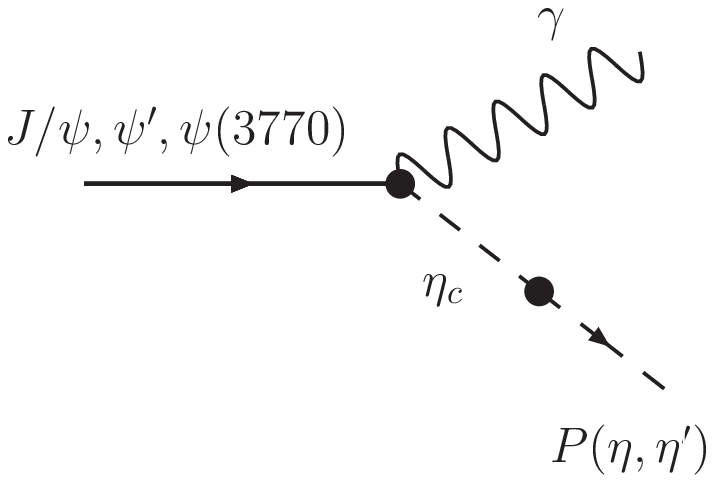}
\caption{ Schematic diagram for $J/\psi (\psi^\prime, \
\psi(3770))\to \gamma \eta$ and $\gamma\eta^\prime$ via
$\eta_c$-$\eta(\eta^\prime)$ mixing. }\label{fig-2}
\end{figure}

Apart from the transitions via Fig.~\ref{fig-1}, another important
transition is via Fig.~\ref{fig-2} which corresponds to the
$\eta_c$-$\eta(\eta^\prime)$ mixing due to the axial vector anomaly.
Note that the process of Fig.~\ref{fig-1}(a) with an intermediate
charmonium does not overlap with Fig.~\ref{fig-2} at the hadronic
level. In fact, it is interesting to note their correlated features:
i) In both cases, the $c\bar{c}$ annihilate at short distances. In
Fig.~\ref{fig-1}(a), the vector configuration of $c\bar{c}$
annihilates into a photon, i.e. $c\bar{c}$ in a relative $S$-wave
with spin-1, while in Fig.~\ref{fig-2} the pseudoscalar $c\bar{c}$
are in a relative $S$-wave but with spin-0, and then annihilates
into gluons. ii) The process of Fig.~\ref{fig-2} is through a
typical magnetic dipole (M1) transition of $J/\psi(\psi^\prime, \
\psi(3770))\to \gamma \eta_c$, which can be regarded as a
non-vector-resonance contribution in respect to the VMD scenario.

With the Lagrangian of Eq.~(\ref{lag-1}), the transition amplitude
can be expressed as
\begin{equation}\label{trans-1}
{\cal M}_{\gamma P}^{VMD}= \left(\sum_V
\frac{1}{p_\gamma^2-M_V^2}\frac{e M_V^2}{f_V}g_{\psi VP}\right){\cal
F}({\bf p}_\gamma^2)
\epsilon_{\mu\nu\alpha\beta}P_\psi^\mu\epsilon_\psi^\nu
p_\gamma^\alpha \epsilon_\gamma^\beta \ ,
\end{equation}
where $g_{\psi VP}$ denotes the coupling constants for the hadronic
vertex of $J/\psi(\psi^\prime, \ \psi(3770))\to VP$, and will be
determined by experimental data via
\begin{equation}\label{trans-2}
{\cal M}_{VP} =g_{\psi VP} {\cal F}({\bf p}_V^2)
\epsilon_{\mu\nu\alpha\beta}P_\psi^\mu\epsilon_\psi^\nu p_V^\alpha
\epsilon_V^\beta \ .
\end{equation}
We adopt an empirical form for the form
factor~\cite{close-amsler,close-kirk,close-zhao-f0}:
\begin{equation}\label{ff-01}
{\cal F}({\bf p}^2)\equiv  e^{-{\bf p}^2/8\beta^2} \ ,
\end{equation}
where parameter $\beta$ is in a range of $300\sim 500$ MeV. This
form factor can be interpreted as the wavefunction overlap which
would be suppressed in a large recoil momentum region for the final
state particles~\cite{close-amsler,close-kirk,close-zhao-f0}. The
incovariant form factor can also be regarded reasonable in this
case. The decay processes are treated in the c.m. frame of the
initial meson. Therefore, the anti-symmetric tensor structure of the
interactions can always be reduced to a form of $M_\psi
\veps_\psi\cdot ({\bf p}_V\times \veps_V)$, which explicitly depends
on the three-vector momentum of the final state vector meson. Note
that for the anti-symmetric tensor couplings all the contributions
to the transition amplitude can be absorbed into the effective
coupling form factor. Because of this, it is natural to expect that
the form factor would contain information of meson wavefunction
overlaps with an explicit three-vector-momentum dependence. In
particular, a harmonic oscillator potential for the quark-antiquark
system will lead to a form factor similar to Eq.~(\ref{ff-01}).

We shall determine the form factor parameter $\beta$ combining the
data for $J/\psi(\psi^\prime)\to VP$ and $\gamma P$. It will then be
fixed and adopted for the calculations of other channels. In the
transition of Fig.~\ref{fig-1}(a), the vector meson will carry the
momentum of the final state photon ${\bf p}_\gamma$.

The transition amplitudes of Fig.~\ref{fig-2} can be expressed as
\begin{eqnarray}
{\cal M}_{\gamma P}^{mixing} &= &\lambda_{P \eta_c}
g_{\psi\gamma\eta_c}
\epsilon_{\mu\nu\alpha\beta}P_\psi^\mu\epsilon_\psi^\nu
p_\gamma^\alpha \epsilon_\gamma^\beta \nonumber\\
&\equiv & \tilde{\lambda}_{P\eta_c} {\cal F}({\bf p}_\gamma^2)
\epsilon_{\mu\nu\alpha\beta}P_\psi^\mu\epsilon_\psi^\nu
p_\gamma^\alpha \epsilon_\gamma^\beta \ ,
\end{eqnarray}
where $\lambda_{P\eta_c}$ is the mixing angle between pseudoscalar
$P$ and $\eta_c$. It has been extracted in Ref.~\cite{Chao:1990im},
$\lambda_{\eta\eta_c}=-4.6\times 10^{-3}$ and
$\lambda_{\eta^\prime\eta_c}=-1.2\times 10^{-2}$, which are also
obtained by Ref.~\cite{Feldmann:1998sh}. It should be noted that in
the above equation the coupling $g_{\psi\gamma\eta_c}$ is extracted
from the data for $J/\psi(\psi^\prime)\to \gamma \eta_c$. The
non-local effects from the off-shell $\eta_c$ at the mass of
$\eta(\eta^\prime)$ have been included in the mixing
angles~\cite{Chao:1990im}. In the second line, we define a reduced
coupling $\tilde{\lambda}_{P\eta_c}\equiv \lambda_{P \eta_c}
g_{\psi\gamma\eta_c}/{\cal F}({\bf p}_\gamma^2)$, which can be
directly compared with the effective coupling $({e}/{f_V}) g_{\psi
VP}$ in Eq.~(\ref{trans-1}).

We do not include the $\eta_c^\prime$ mixings with the
$\eta(\eta^\prime)$ in $J/\psi\to\gamma \eta$ and
$\gamma\eta^\prime$ since their mixing angles are relatively small.
Nevertheless, the $\eta_c^\prime$ mixing effects will be further
suppressed by the unknown but believe-to-be-small branching ratio
for $\eta_c^\prime\to J/\psi\gamma$.

\section{Numerical results}

\subsection{Results from VMD}

In Table~\ref{tab-1}, the data for $J/\psi$, $\psi^\prime$ and
$\psi(3770)\to VP$ from PDG 2010~\cite{Nakamura:2010zzi} are listed.
It shows that most of the light $VP$ channels have been measured for
$J/\psi$ and $\psi^\prime$ hadronic decays. In contrast, most of the
light $VP$ channels for $\psi(3770)$ are below the experimental
precision limit except for $\phi\eta$. As mentioned earlier, the
$J/\psi(\psi^\prime)\to VP$ channels are correlated with the
so-called ``$\rho\pi$ puzzle" in the literature. However, our
attention in the present work is different. We shall use the
experimental data for $J/\psi(\psi^\prime, \ \psi(3770))\to VP$ as
an input to investigate the role played by the VMD mechanisms in the
vector charmonium radiative decays. This treatment means that one
need not be concerned about the detailed transition mechanisms for
$J/\psi(\psi^\prime, \ \psi(3770))\to VP$ at this moment since they
all have been contained in the experimental data. We emphasize that
this should not be a trivial starting point. Success of such a
prescription would help us clarify two major processes in the
charmonium radiative decays, i.e. the relative $S$-wave $c\bar{c}$
annihilations would occur either via spin-1 or spin-0
configurations.

\begin{center}
\begin{table}
\begin{tabular}{|c|c|c|c|}
 \hline
  \hline
  Channels  & $J/\psi$ & $\psi^\prime$ & $\psi(3770)$ \\
  \hline
  $J/\psi\pi^0$ & - &  $(1.30\pm 0.10)\times 10^{-3}$ & $<2.8\times 10^{-4}$ \\
  $J/\psi\eta$ & - &  $(3.28\pm 0.07)\%$  & $(9\pm 4)\times 10^{-4}$ \\
  \hline
  $\rho\pi$ & $(1.69\pm 0.15)\%$ &  $(3.2\pm 1.2)\times 10^{-5}$ & - \\
  $\rho^0\pi^0$ &  $(5.6\pm 0.7)\times 10^{-3}$ & $\dots$ & - \\
  $\omega\pi^0$  &  $(4.5\pm 0.5)\times 10^{-4}$  &  $(2.1\pm 0.6)\times 10^{-5}$  & - \\
  $\phi\pi^0$   &  $<6.4\times 10^{-6}$  &  $< 4\times 10^{-6}$  & - \\
  $\omega\eta$ & $(1.74\pm 0.20)\times 10^{-3}$ &  $<1.1\times 10^{-5}$   & - \\
  $\phi\eta$ & $(7.5\pm 0.8)\times 10^{-4}$ &  $(2.8^{+1.0}_{-0.8})\times 10^{-5}$ & $(3.1\pm 0.7)\times 10^{-4}$ \\
  $\rho\eta$  & $(1.93\pm 0.23)\times 10^{-4}$ &  $(2.2\pm 0.6)\times 10^{-5}$  & - \\
  $\omega\eta^\prime$ &  $(1.82\pm 0.21)\times 10^{-4}$ &  $(3.2^{+2.5}_{-2.1})\times 10^{-5}$   & - \\
  $\phi\eta^\prime$  &  $(4.0\pm 0.7)\times 10^{-4}$ &  $(3.1\pm 1.6)\times 10^{-5}$   & - \\
  $\rho\eta^\prime$ &  $(1.05\pm 0.18)\times 10^{-4}$ &  $(1.9^{+1.7}_{-1.2})\times 10^{-5}$  & -\\
  \hline
\end{tabular}
\caption{Branching ratios for $J/\psi(\psi^\prime, \ \psi(3770))\to
VP$ from PDG 2010~\protect\cite{Nakamura:2010zzi}. The dash ``-" and
dots ``$\dots$" denote the forbidden and unavailable channels,
respectively. } \label{tab-1}
\end{table}
\end{center}

\begin{center}
\begin{table}
\begin{tabular}{|c|c|c|c|}
 \hline
  \hline
  $VP$  & $(e/f_V)g_{J/\psi VP}$ & $(e/f_V)g_{\psi^\prime VP}$  & $(e/f_V)g_{\psi(3770) VP}$ \\
  \hline
  $J/\psi\pi^0$ & - &  $4.02\times 10^{-4}$ & $<9.73\times 10^{-4}$ \\
  $J/\psi\eta$ & - &  $6.25\times 10^{-3}$  & $2.74\times 10^{-3}$ \\
  $J/\psi\eta^\prime$ & - & $3.01\times 10^{-2}$ &  $1.32\times 10^{-2}$ \\
  $\psi^\prime\pi^0$ & $2.40\times 10^{-4}$ &  -  & - \\
  $\psi^\prime\eta$ & $3.74\times 10^{-3}$  &  -  & - \\
  $\psi^\prime\eta^\prime$ & $1.80\times 10^{-2}$ & - & - \\
  $\psi(3770)\pi^0$ & $<3.22\times 10^{-4}$ & - & - \\
  $\psi(3770)\eta$  & $9.09\times 10^{-4}$ & - & - \\
  $\psi(3770)\eta^\prime$ & $4.37\times 10^{-3}$ & - & - \\
  \hline
  $\rho^0\pi^0$ & $2.83\times 10^{-3}$ &  $6.69\times 10^{-4}$ & $\dots$ \\
  $\omega\pi^0$  &  $2.35\times 10^{-4}$  &  $2.73\times 10^{-4}$  & $\dots$ \\
  $\phi\pi^0$   &  $<2.51\times 10^{-5}$  &  $< 1.02\times 10^{-4}$  & $\dots$ \\
  $\omega\eta$ & $3.97\times 10^{-4}$ &  $<1.67\times 10^{-4}$   & $\dots$ \\
  $\phi\eta$ & $2.72\times 10^{-4}$ &  $2.69\times 10^{-4}$  & $1.02\times 10^{-2}$  \\
  $\rho\eta$  & $4.52\times 10^{-4}$ &  $8.10\times 10^{-4}$  & $\dots$\\
  $\omega\eta^\prime$ &  $9.54\times 10^{-5}$ &  $2.02\times 10^{-4}$   & $\dots$ \\
  $\phi\eta^\prime$  &  $1.48\times 10^{-4}$ &  $1.20\times 10^{-4}$   & $\dots$ \\
  $\rho\eta^\prime$ &  $2.48\times 10^{-4}$ &  $5.34\times 10^{-4}$  & $\dots$ \\
  \hline
\end{tabular}
\caption{Effective couplings $\frac{e}{f_V} g_{\psi VP}$ (in unit of
GeV$^{-1}$) for $J/\psi(\psi^\prime, \ \psi(3770))\to \gamma P$
extracted from the intermediate $VP$ channels. Note that the form
factor ${\cal F}({\bf p}_\gamma^2)$ is not included. The dash ``-"
and dots ``$\dots$" denote the forbidden and unavailable channels,
respectively. } \label{tab-2}
\end{table}
\end{center}

In Table~\ref{tab-1}, the branching ratios for $\psi^\prime$ and
$\psi(3770)\to J/\psi\eta$ and $J/\psi\pi^0$ are also listed. As
pointed out earlier, these channels are rather important for
understanding the observed branching ratio patterns. The effective
coupling $g_{\psi VP}$ in Eqs.~(\ref{trans-1}) and (\ref{trans-2})
is a scale-independent constant. The data in Table~\ref{tab-1} will
allow us to extract $g_{\psi VP}$ for different $VP$ channels in
association with the form factor parameter $\beta$. The overall
numerical study suggests that a smaller value of $\beta=0.3$ GeV is
favored. This is due to that in $J/\psi(\psi^\prime)\to \gamma P$,
the intermediate vector mesons are in a highly virtual kinematic
region. Part of the off-shell effects would be absorbed into the
form factor parameter $\beta$ as we adopt the $V\to \gamma^*$
couplings $e/f_V$ which are determined by data for $V\to e^+
e^-$~\cite{Nakamura:2010zzi}.

In Table~\ref{tab-2} we list the joint coupling constants
$(e/f_V)g_{\psi VP}$ for different $VP$ channels as a reference.
These quantities are the corresponding scale-independent couplings
in $J/\psi(\psi^\prime)\to \gamma P$, and provide an immediate
estimate of the relative strengths among those transitions
amplitudes that involve different vector mesons. The form factor
${\cal F}({\bf p}_\gamma^2)=e^{-{\bf p}_\gamma^2/8\beta^2}$ with
$\beta=0.3$ GeV will lead to an overall suppression to the vertices.
In the light $VP$ sector, the strong $\rho^0\gamma$ coupling
accounts for the relatively large contributions from the $\rho^0$
mediated transitions.

In the vector-charmonium-mediated channels, the non-negligible
coupling of $g_{\psi^\prime J/\psi\eta}$ implies a non-vanishing
coupling of $g_{\psi^\prime J/\psi\eta^\prime}$, although the decay
of $\psi^\prime\to J/\psi \eta^\prime$ is prohibited by the phase
space. The influence of $\psi^\prime\to J/\psi \eta^\prime$ in
$\psi^\prime\to \gamma \eta^\prime$ should not be neglected and must
be included in the amplitude. As we know, the $\eta$ and
$\eta^\prime$ can be expressed as mixtures of quark flavor singlets:
\begin{eqnarray}\label{mix-eta-etap}
\eta & = & \cos\alpha_P n\bar{n} -\sin\alpha_P s\bar{s} \ ,
\nonumber\\
\eta^\prime & = & \sin\alpha_P n\bar{n} + \cos\alpha_P s\bar{s} \ ,
\end{eqnarray}
where $\alpha_P\equiv \mbox{arctan}\sqrt{2} +\theta_P$, and
$\theta_P\simeq -11.7^\circ$ is the SU(3) flavor singlet and octet
mixing angle. Thus, we have
\begin{equation}\label{etap-coup}
g_{\psi^\prime J/\psi\eta^\prime} = g_{\psi^\prime J/\psi\eta}\left(
\frac{\sqrt{2}\sin\alpha_P+R\cos\alpha_P}{\sqrt{2}\cos\alpha_P-R\sin\alpha_P}\right)
\ ,
\end{equation}
where $R$ describes the SU(3) flavor symmetry breaking. In general,
$R\equiv f_\pi/f_K \simeq 0.838$ is commonly adopted for the
relative production strength of an $s\bar{s}$ to $q\bar{q}$. The
above relation is based on the $q\bar{q}$ and $s\bar{s}$ mixing
scheme~\cite{Li:2007ky,Thomas:2007uy,Escribano:2007cd,Cheng:2008ss,Mathieu:2010ss}
and does not include a possible glueball component. If one extends
the $\eta$-$\eta^\prime$ mixing to accommodate the glueball ${\cal
G}$, the coupling of $g_{\psi^\prime J/\psi\eta^\prime}$ can be
expressed as
\begin{equation}\label{etap-coup-2}
g_{\psi^\prime J/\psi\eta^\prime} = g_{\psi^\prime J/\psi\eta}
\left(\frac{\sqrt{2}X_{\eta^\prime}+R Y_{\eta^\prime}+ G
Z_{\eta^\prime}}{\sqrt{2}X_\eta +R Y_\eta + G Z_\eta}\right) \ ,
\end{equation}
where parameter $G$ denotes the relative strength of producing the
pseudoscalar glueball ${\cal G}$ to a light $q\bar{q}$ component.
The general flavor wavefunctions for $\eta$ and $\eta^\prime$ are
\begin{eqnarray}
\eta & = & X_\eta n\bar{n} + Y_\eta s\bar{s} + Z_\eta {\cal G} \
,\nonumber\\
\eta^\prime & =& X_{\eta^\prime} n\bar{n} + Y_{\eta^\prime} s\bar{s}
+ Z_{\eta^\prime}{\cal G} \ ,
\end{eqnarray}
for which different model solutions can be found in the
literature~\cite{Li:2007ky,Thomas:2007uy,Escribano:2007cd,Cheng:2008ss,Seiden:1988rr}.
Generally speaking, the introduction of the glueball component will
introduce new parameters. Taking into account that the glueball
components within the $\eta$ and $\eta^\prime$ are rather small, and
Eq.~(\ref{mix-eta-etap}) is well established to leading accuracy, we
neglect the possible glueball mixing effects in the present
analysis.

We adopt the same on-shell couplings of $g_{J/\psi\psi^\prime P}$ as
those extracted in $\psi^\prime\to J/\psi P$ since the kinematics
for these two processes are similar to each other. Namely, we
neglect the off-shell effects with the couplings of
$g_{J/\psi\psi^\prime P}$ in contrast with $g_{\psi^\prime J/\psi
P}$.

As listed in Table~\ref{tab-2}, it shows that the charmonium poles
are one of the most important contributing sources to the
$J/\psi(\psi^\prime)\to \gamma \eta$ and $\gamma\eta^\prime$, which
seems to be slightly out of expectation and has not been addressed
before. This feature is explicit for the $\psi^\prime$ decays since
the decay of $\psi^\prime\to J/\psi\eta$ is experimentally
accessible. In contrast, other $VP$ channels' contributions to
$\gamma P$ are rather small due to their relatively small branching
ratios. Similar phenomena appear in $J/\psi\to VP$ except that the
sizeable branching ratio for $J/\psi\to \rho\pi$ would also make the
$\rho\pi$ channel an important contributor to the $\gamma P$
amplitude.

\subsection{Results from $\eta_c$-$\eta(\eta^\prime)$ mixings}

In Table~\ref{tab-3}, we list the effective couplings derived from
the $\eta_c$-$\eta(\eta^\prime)$ mixings~\cite{Chao:1990im}. These
values can be directly compared with $(e/f_V)g_{\psi VP}$ listed in
Table~\ref{tab-2}. It shows that in $J/\psi\to \gamma \eta$ and
$\gamma\eta^\prime$, the axial-anomaly-driving mixing contributions
turn out to be more predominant than the VMD, while in
$\psi^\prime\to \gamma P$ the most important contribution is from
the $J/\psi$ pole.

We list the individual branching ratios given by the VMD and
$\eta_c$-$\eta(\eta^\prime)$ mixings in Table~\ref{tab-4} as a
comparison. Indeed, it shows that the mixing contributions have
nearly saturated the branching ratios in $J/\psi\to \gamma \eta$ and
$\gamma\eta^\prime$. However, the situation changes in $\psi^\prime$
decays where the VMD mechanisms become more important. An
interesting feature is that one in principle needs both to give an
overall account of the measured branching ratios.

Note that in Table~\ref{tab-4}, the ranges of uncertainties for the
VMD results are given by the experimental errors in
Table~\ref{tab-1}.

\begin{center}
\begin{table}
\begin{tabular}{|c|c|c|}
 \hline
  \hline
    & $\tilde{\lambda}_{P\eta_c}(J/\psi \to \gamma P)$ & $\tilde{\lambda}_{P\eta_c}(\psi^\prime \to \gamma P)$   \\
  \hline
$\gamma \eta$         &  $2.10\times 10^{-2}$    &  $5.12\times 10^{-3}$   \\
$\gamma \eta^\prime$  &  $3.66\times 10^{-2}$    &  $8.91\times 10^{-3}$   \\
  \hline
\end{tabular}
\caption{Reduced effective couplings (in unit of GeV$^{-1}$) from
the $\eta_c$-$\eta(\eta^\prime)$ mixings.  } \label{tab-3}
\end{table}
\end{center}

\begin{center}
\begin{table}
\begin{tabular}{|c|c|c|c|c|}
 \hline
  \hline
  & \multicolumn{2}{c|}{$J/\psi\to \gamma P$} & \multicolumn{2}{c|}{$\psi^\prime \to \gamma P$}   \\
  \hline
  $\gamma P$  & VMD & $\eta_c$ mixing & VMD & $\eta_c$ mixing   \\
  \hline
$\gamma \pi^0$ & $(1.64\sim 2.04)\times 10^{-5}$ & - & $(0.66\sim
1.15) \times 10^{-7}$ & - \\
$\gamma\eta$ & $(0.060\sim 0.063)\times 10^{-3}$ &  $0.61\times
10^{-3}$ & $(3.33\sim 3.61)\times 10^{-6}$ & $1.62\times 10^{-6}$ \\
$\gamma\eta^\prime$ &  $(1.04\sim 1.05)\times 10^{-3}$ & $3.50\times
10^{-3}$ & $(0.58\sim 0.61)\times 10^{-4}$ & $0.096\times 10^{-4}$
\\
  \hline
\end{tabular}
\caption{Branching ratios for $J/\psi(\psi^\prime)\to \gamma \eta$
and $\gamma\eta^\prime$ given by the VMD mechanisms and
$\eta_c$-$\eta(\eta^\prime)$ mixings, respectively. } \label{tab-4}
\end{table}
\end{center}

\subsection{Discussions}

To compare with the experimental measurements, we need to add the
VMD and $\eta_c$-$\eta(\eta^\prime)$ mixing amplitudes to each other
coherently. Taking the advantage of the unique Lorentz structure of
the $VVP$ coupling, we can express the total transition amplitude as
follows,
\begin{equation}
{\cal M}_{tot} = {\cal M}_{\gamma P}^{VMD}+e^{i\delta}{\cal
M}_{\gamma P}^{mixing} \ ,
\end{equation}
where $\delta$ is introduced to take into account possible phase
differences between these two amplitudes. In the transition
processes that we are interested in here, such a phase ambiguity
seems inevitable due to a important role played by hadronic
transition mechanisms. Since several different hadronic level
amplitudes are involved in ${\cal M}_{\gamma P}^{VMD}$, it is not a
necessity that ${\cal M}_{\gamma P}^{VMD}$ and ${\cal M}_{\gamma
P}^{mixing}$ should share the same phase angle for different
pseudoscalar channels. We expect that the experimental
data~\cite{Nakamura:2010zzi,bes-iii} would provide a constraint on
it.

\begin{figure}
\begin{tabular}{cc}
  \includegraphics[scale=0.3]{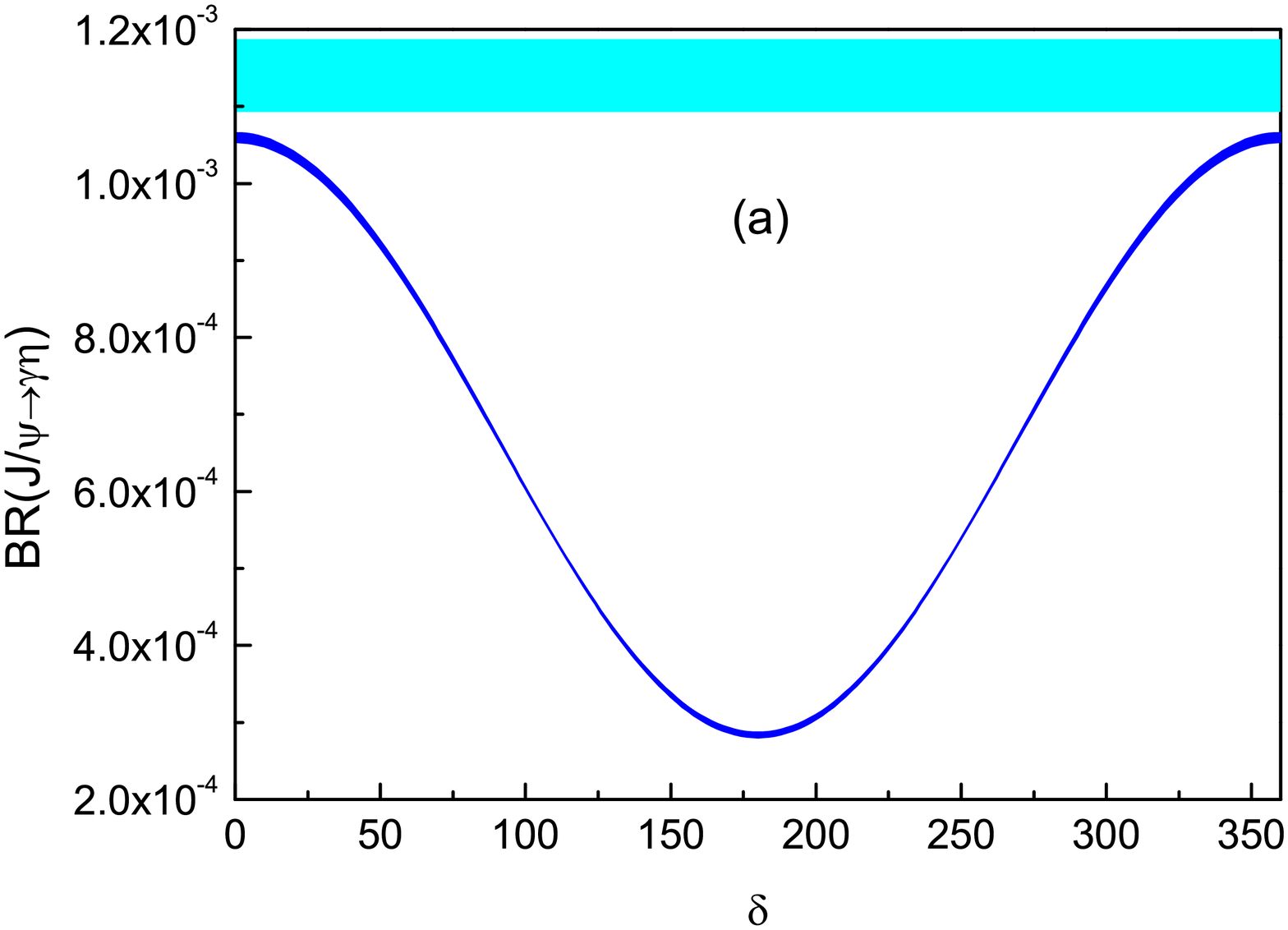} & \includegraphics[scale=0.3]{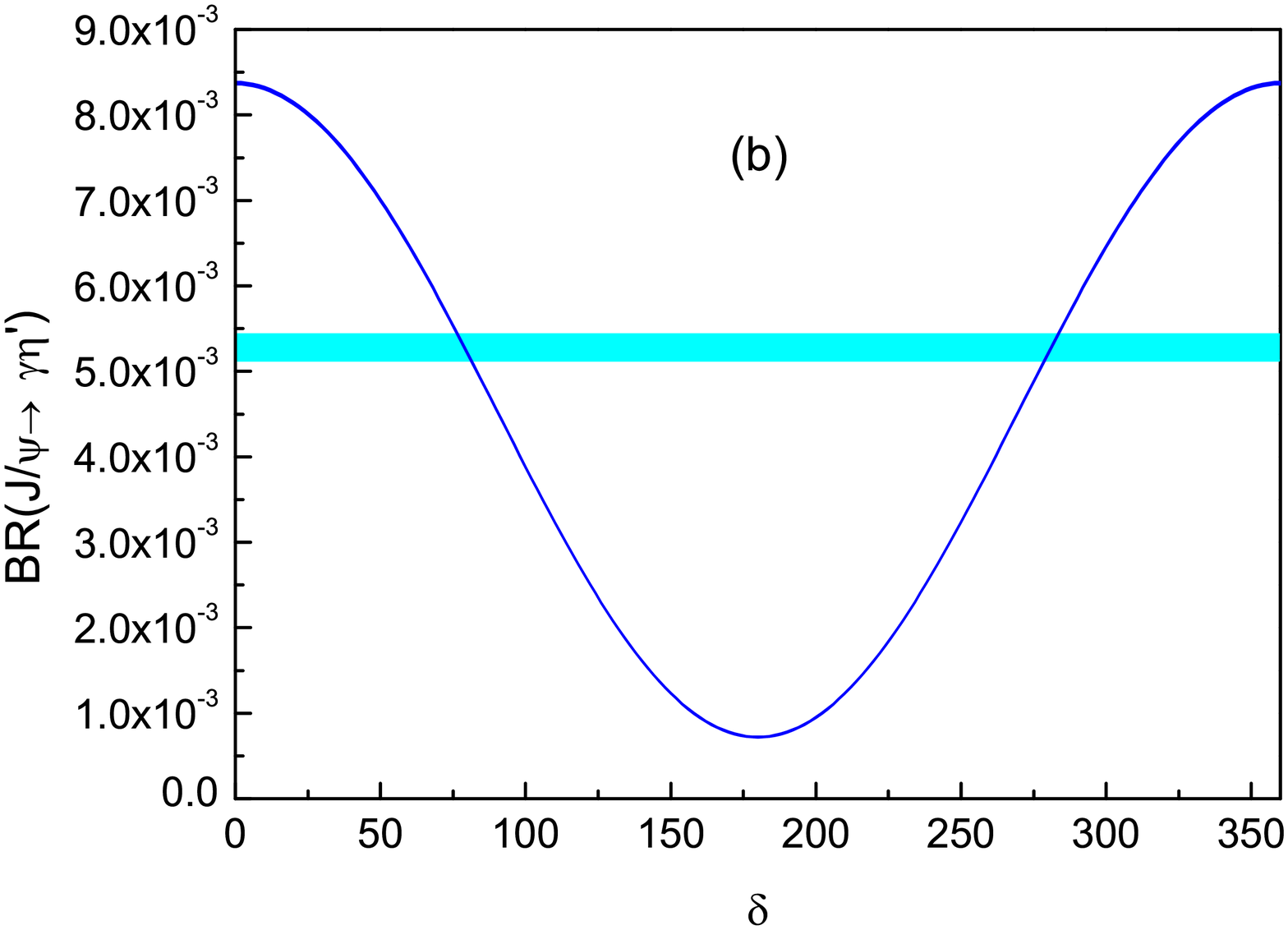} \\
  \includegraphics[scale=0.3]{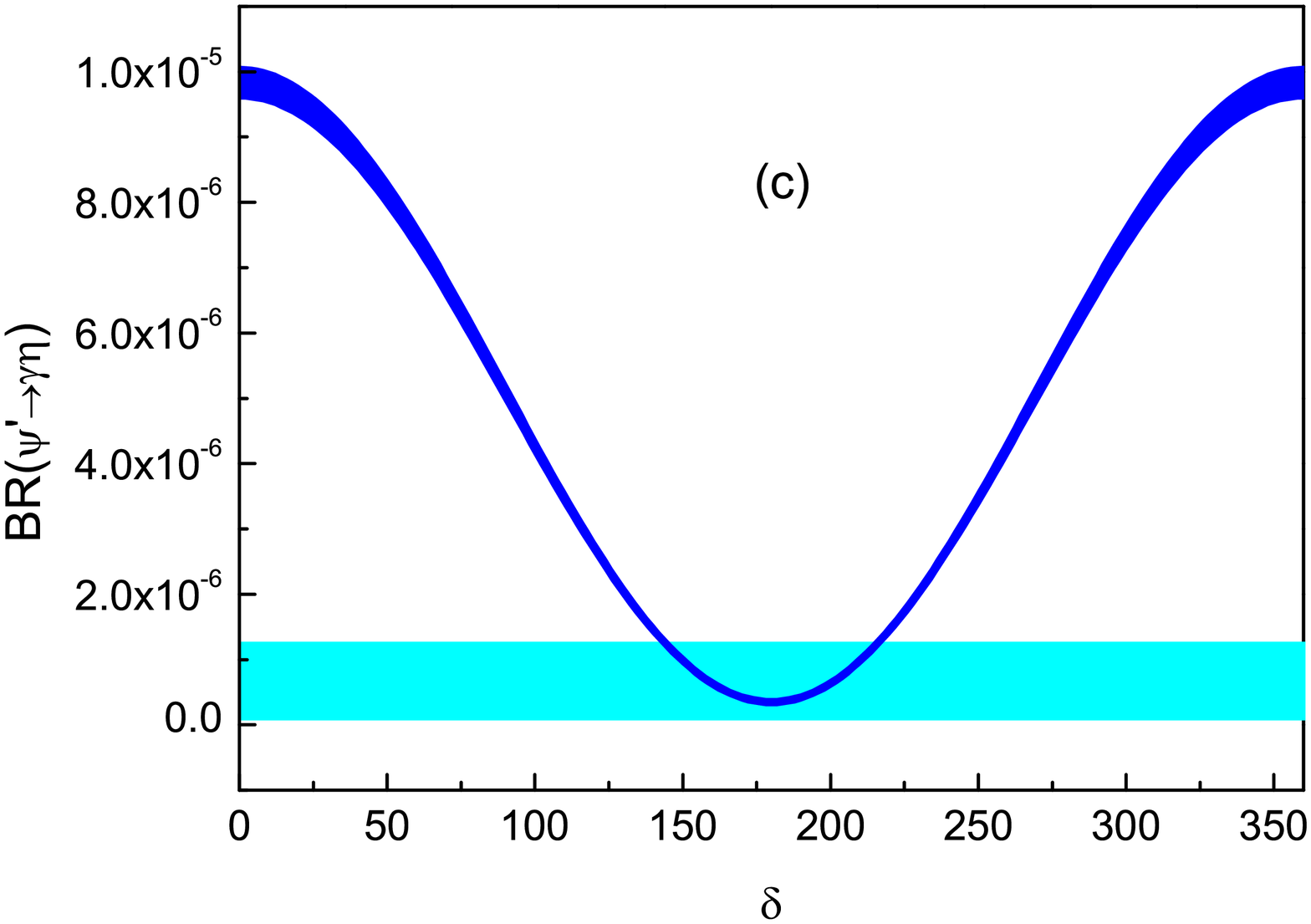} & \includegraphics[scale=0.3]{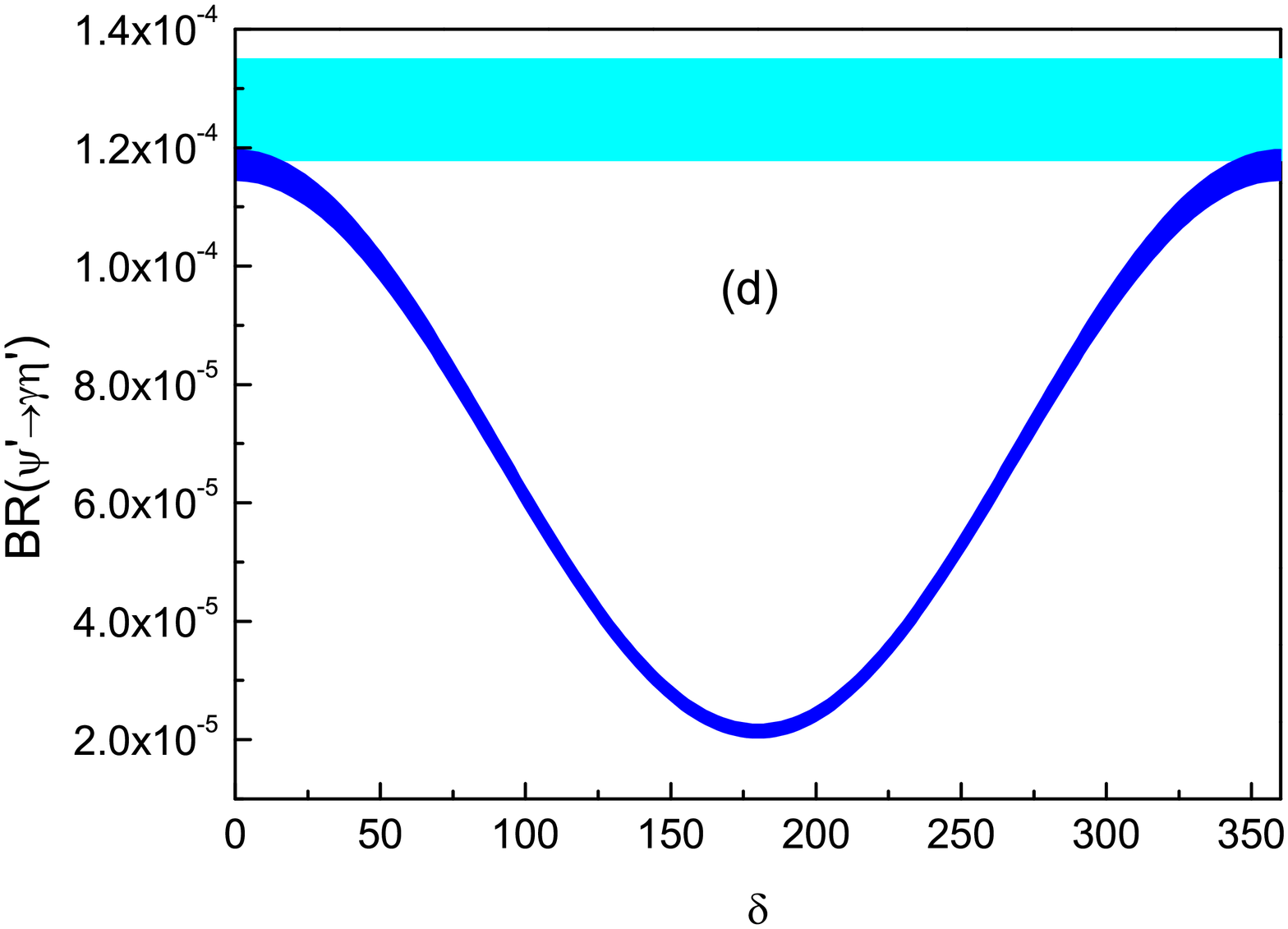}\\
\end{tabular}
\caption{The phase angle $\delta$-dependence of branching ratios for
$J/\psi\to \gamma\eta^{(\prime)}$ ((a) and (b)) and $\psi^\prime\to
\gamma \eta^{(\prime)}$ ((c) and (d)). The experimental data with
uncertainties are shown as the straight bands, while the theoretical
results are shown by the curvilinear bands. The theory uncertainties
are given by the uncertainties of the data for
$J/\psi(\psi^\prime)\to VP$ as listed in Table~\protect\ref{tab-1}.
}\label{fig-3}
\end{figure}

In Fig.~\ref{fig-3}, we plot the $\delta$-dependence of the
branching ratios in comparison with the PDG2010
averages~\cite{Nakamura:2010zzi} and new experimental data from
BESIII~\cite{bes-iii}. It shows that in the two decays, $J/\psi\to
\gamma\eta$ and $\psi^\prime\to \gamma\eta^\prime$, the transition
amplitudes of the VMD and $\eta_c$-$\eta(\eta^\prime)$ mixings are
well in phase. In contrast, they seem to be out of phase in
$\psi^\prime\to \gamma\eta$, although the experimental uncertainties
are quite large. The central value of the data can be best accounted
for at $\delta\simeq 140^\circ$ or $220^\circ$. More complex phases
appear in $J/\psi\to\gamma \eta^\prime$, although the dominant
contributions are from the axial gluonic anomaly. In this case, the
phase angle $\delta=80^\circ$ or $280^\circ$ are favored. It should
be mentioned that in a recent paper by
BESIII~\cite{Collaboration:2010kp}, a smaller branching ratio for
$J/\psi\to\gamma\eta^\prime$ is reported, i.e. $BR(J/\psi\to
\gamma\eta^\prime)=(4.86\pm 0.03\pm 0.24)\times 10^{-3}$. This value
is consistent with the PDG2010 average, and would favor
$\delta\simeq 90^\circ$ or $270^\circ$.

In Table~\ref{tab-5}, we list the coherent results for the branching
ratios $BR(J/\psi\to \gamma P)$ and $BR(\psi^\prime\to \gamma P)$ in
comparison with the data again~\cite{Nakamura:2010zzi,bes-iii}. The
phase angles are fixed as shown in Fig.~\ref{fig-3} with the best
description of the central value of the data. Again, the theoretical
uncertainties due to adopting the data for $J/\psi(\psi^\prime)\to
VP$ are included. We also include the $\psi(3770)\to \gamma P$ as a
prediction of the VMD mechanism. The predicted branching ratios are
all small. The $\eta_c$ mixing contributions are not included here
due to lack of data. Also,  most of the light vector meson
contributions to the $\psi(3770)$ radiative decays are rather small
and unavailable. Thus, the predicted branching ratios are actually
given by the $J/\psi$ pole in the VMD model.  Given the same
statistics for $\psi(3770)$ as the $\psi^\prime$ from BESIII, the
accessible channel would be $\psi(3770)\to \gamma \eta^\prime$.
Experimental examination of the predicted pattern in
Table~\ref{tab-5} would be an interesting test of the VMD mechanisms
proposed in this work.

In general, the results fit the observed branching ratio pattern
very well, except that the branching ratio for $\psi^\prime \to
\gamma\pi^0$ seems to have some discrepancies. It might be a sign
that other non-VMD mechanisms may also play a role. For $J/\psi\to
\gamma\pi^0$, the dominance of $J/\psi\to \rho^0\pi^0$ can naturally
account for the data. It should be mentioned that
Ref.~\cite{Rosner:2009bp} also confirms the VMD contributions via
the $\rho^0\pi^0$ channel to $J/\psi\to \gamma\pi^0$.

\begin{center}
\begin{table}
\begin{tabular}{|c|c|c|c|c|c|c|}
 \hline
  \hline
  & \multicolumn{2}{c|}{$J/\psi\to \gamma P$} & \multicolumn{2}{c|}{$\psi^\prime \to \gamma P$} & \multicolumn{2}{c|}{$\psi(3770) \to \gamma P$}  \\
  \hline
  $\gamma P$  & Experiment & Theory & Experiment & Theory & Experiment & Theory \\
  \hline
  $\gamma\pi^0$ &  $(3.49^{+0.33}_{-0.30})\times 10^{-5}$ & $(1.64\sim 2.04)\times 10^{-5}$  &  $(1.58\pm 0.40\pm 0.13)\times
  10^{-6}$ & $(0.66\sim 1.15)\times
  10^{-7}$ &  $<2\times 10^{-4}$  &    $3.25\times 10^{-9}$      \\
  $\gamma\eta$  &  $(1.104\pm 0.034)\times 10^{-3}$  &  $(1.05\sim 1.06)\times 10^{-3}$  &  $(1.38\pm 0.48\pm 0.09)\times
  10^{-6}$  &  $(1.39 \sim 1.53)\times
  10^{-6} $  &  $<1.5\times 10^{-4}$  &   $7.95\times 10^{-7}$     \\
  $\gamma\eta^\prime$  &  $(5.28\pm 0.15)\times 10^{-3}$  &  $5.20\sim 5.22\times 10^{-3}$  & $(1.26\pm 0.03\pm 0.08)\times
  10^{-4}$ &  $(1.14\sim 1.19)\times
  10^{-4}$  &  $<1.8\times 10^{-4}$  &   $1.64\times 10^{-6}$      \\
  \hline
\end{tabular}
\caption{Calculated branching ratios for $J/\psi (\psi^\prime, \
\psi(3770))\to \gamma P$ based on the VMD model. Experimental data
from PDG 2010~\protect\cite{Nakamura:2010zzi} for $J/\psi$ and
$\psi(3770)$ decays and from BESIII~\protect\cite{bes-iii} for
$\psi^\prime$ decays are included as a comparison. The phase angles
are fixed in such a way that the theoretical results can best
describe the central values of the experimental data.} \label{tab-5}
\end{table}
\end{center}

Our investigation suggests the importance of a coherent treatment
for the VMD mechanism and $\eta_c$-$\eta(\eta^\prime)$ mixings. Note
that the charmonium pole contribution has not been included by the
previous
studies~\cite{Novikov:1979uy,Chao:1990im,Feldmann:1998sh,ali,petrov,Seiden:1988rr,Thomas:2007uy}.
Meanwhile, an understanding of why the VMD and axial gluonic anomaly
mechanisms play different roles in $J/\psi$ and $\psi^\prime$ decays
would be essentially important. The following points may help to
clarify this question:

i) As mentioned earlier, there are some interesting correspondences
between the axial gluonic anomaly and VMD in this case. In the axial
gluonic anomaly transitions the $c\bar{c}$ annihilate into gluon
fields at short distances in a relative $S$-wave and spin-0, which
induces mixings with the Goldstone boson $\eta$ and SU(3) flavor
singlet $\eta^\prime$. The photon radiation can be regarded as from
non-vector-resonance M1 transitions. In the VMD transitions via the
charmonium state, the $c\bar{c}$ also annihilate at short distances
in a relative $S$-wave, but with spin-1. In this case, the
annihilated $c\bar{c}$ couple to a photon, and radiate two soft
gluons which can couple to pseudoscalar states.

ii) The difference between those two mechanisms can be
well-understood quantum mechanically. For $J/\psi\to \gamma
\eta^{(\prime)}$, the VMD transitions via $\psi^\prime$ pole is
relative suppressed by the $\psi^\prime\gamma$ coupling since as the
first radial excited state the wavefunction of $\psi^\prime$ at the
origin is smaller than that of $J/\psi$. In contrast, the
axial-gluonic-anomaly-driving $\eta_c$-$\eta(\eta^\prime)$ mixings
will occur via $J/\psi\to \gamma\eta_c\to \gamma\eta^{(\prime)}$,
where the first step is a typical EM M1 transition between two
ground charmonium states. It is allowed by the quantum transition
selection rule at leading order.

The situation changes in $\psi^\prime\to \gamma\eta^{(\prime)}$. On
the one hand, the VMD transition will be dominated by the $J/\psi$
pole, which will be coupled to the EM field. On the other hand, the
axial gluonic anomaly transitions via the $\psi^\prime$-$\eta_c$ M1
transition will be suppressed by the quantum transition selection
rule at leading order. For the $\eta_c^\prime$-mediated transition,
the $\eta_c^\prime$ mixings with the $\eta$ and $\eta^\prime$ will
then be suppressed~\cite{Chao:1990im}.

The above qualitative argument explains why the VMD mechanism and
axial gluonic anomaly play different roles in $J/\psi$ and
$\psi^\prime$ decays, respectively, as manifested by the
calculation. In particular, it shows that both mechanisms are
crucial for our understanding of the observed branching ratio
patterns.

The successful account of the observed branching ratio patterns for
$J/\psi(\psi^\prime)\to \gamma P$ in the VMD model has an important
implication of the hadronic decay mechanisms for
$J/\psi(\psi^\prime)\to VP$. It shows that the ``puzzling" radiative
decay patterns in $J/\psi(\psi^\prime)\to \gamma P$ have direct
connections with the hadronic decay mechanisms, i.e.
$J/\psi(\psi^\prime)\to VP$, instead of some other abnormal
processes. As a consequence, it will guide our further
investigations of the transitions of $J/\psi(\psi^\prime)\to VP$,
and impose constraints on processes such as illustrated by
Fig.~\ref{fig-1}. For instance, the hadronic part of
Fig.~\ref{fig-1}(c) is found to be an important non-perturbative QCD
mechanism that contributes predominantly in $\psi^\prime\to
J/\psi\eta$ and $J/\psi\pi^0$~\cite{Guo:2010ak,Guo:2009wr}. As
pointed out recently in a series of papers on the subject of
non-perturbative transition mechanisms in charmonium
decays~\cite{Zhang:2009kr,Zhao:2008eg,Liu:2009vv,Liu:2010um,Wang:2010iq,Guo:2010ak,Guo:2009wr},
such intermediate meson loop transitions would be an natural
mechanism for evading the pQCD helicity selection rule and
explaining the ``$\rho\pi$ puzzle" in $J/\psi(\psi^\prime)\to VP$.

\section{Summary}

In brief, with the available data for $J/\psi(\psi^\prime)\to VP$,
we show that the VMD model is still useful for our understanding of
the newly measured branching ratios for $J/\psi(\psi^\prime)\to
\gamma P$ in association with the $\eta_c$-$\eta(\eta^\prime)$
mixings via the axial gluonic anomaly. Importance of such a
contribution has not been recognized before. In particular, we
stress that the intermediate vector charmonia can have significant
contributions via e.g. $\psi^\prime\to J/\psi\eta \to \gamma\eta$.
We show that these two mechanisms behave differently in $J/\psi$ and
$\psi^\prime\to \gamma P$, and can be understood by state transition
selection rules. We also emphasize that the consistency between
$J/\psi(\psi^\prime)\to \gamma P$ and $VP$ demonstrated in this work
would impose important constraints on the non-pQCD mechanisms in
$J/\psi(\psi^\prime)\to VP$. It would be useful for our final
understanding of the long-standing ``$\rho\pi$ puzzle" in
$J/\psi(\psi^\prime)\to VP$.

\section*{Acknowledgement}

The author thanks useful discussions with H.-N. Li, X.-Q. Li, and
H.-W. Ke. This work is supported, in part, by the National Natural
Science Foundation of China (Grants No. 10491306), Chinese Academy
of Sciences (KJCX2-EW-N01), and Ministry of Science and Technology
of China (2009CB825200).


\begin{thebibliography}{99}

\bibitem{:2009tia}
  T.~K.~Pedlar {\it et al.}  [CLEO Collaboration],
  Phys.\ Rev.\  D {\bf 79}, 111101 (2009)
  [arXiv:0904.1394 [hep-ex]].

\bibitem{Amsler:2008zzb}
  C.~Amsler {\it et al.}  [Particle Data Group],
  Phys.\ Lett.\  B {\bf 667}, 1 (2008).

\bibitem{bes-iii} M. Ablikim {\it et al.} [BESIII Collaboration], Phys. Rev. Lett. {\bf 105}, 261801 (2010) [arXiv:1011.0889[hep-ex]];
L.L. Wang (for BESIII Collaboration), talk given at The 4th
International Workshop on Charm Physics - Charm 2010, 2010, Beijing.

\bibitem{Nakamura:2010zzi}
  K.~Nakamura {\it et al.}  [Particle Data Group],
  J.\ Phys.\ G {\bf 37}, 075021 (2010).

\bibitem{Novikov:1979uy}
  V.~A.~Novikov, M.~A.~Shifman, A.~I.~Vainshtein and V.~I.~Zakharov,
  Nucl.\ Phys.\  B {\bf 165}, 55 (1980).

\bibitem{Chao:1990im}
  K.~T.~Chao,
  Nucl.\ Phys.\  B {\bf 335} (1990) 101.

\bibitem{Feldmann:1998sh}
  T.~Feldmann, P.~Kroll and B.~Stech,
  Phys.\ Lett.\  B {\bf 449}, 339 (1999)
  [arXiv:hep-ph/9812269].



\bibitem{ali} A. Ali, J. Chay, C. Greub, and P. Ko, Phys. Lett. B
{\bf 424}, 161 (1998).

\bibitem{petrov} A. Petrov, Phys. Rev. D {\bf 58}, 054004 (1998).

\bibitem{Bauer:1977iq}
  T.~H.~Bauer, R.~D.~Spital, D.~R.~Yennie and F.~M.~Pipkin,
  Rev.\ Mod.\ Phys.\  {\bf 50}, 261 (1978)
  [Erratum-ibid.\  {\bf 51}, 407 (1979)].

\bibitem{Bauer:1975bw}
  T.~Bauer and D.~R.~Yennie,
  Phys.\ Lett.\  B {\bf 60}, 169 (1976).

\bibitem{Li:2007ky}
  G.~Li, Q.~Zhao and C.~H.~Chang,
  J.\ Phys.\ G {\bf 35}, 055002 (2008)
  [arXiv:hep-ph/0701020].

\bibitem{Zhao:2006gw}
  Q.~Zhao, G.~Li and C.~H.~Chang,
  Phys.\ Lett.\  B {\bf 645}, 173 (2007)
  [arXiv:hep-ph/0610223].

\bibitem{Zhang:2009kr}
  Y.~J.~Zhang, G.~Li and Q.~Zhao,
  Phys.\ Rev.\ Lett.\  {\bf 102}, 172001 (2009)
  [arXiv:0902.1300 [hep-ph]].

\bibitem{Zhao:2010ja}
  Q.~Zhao,
  Nucl.\ Phys.\ Proc.\ Suppl.\  {\bf 207-208}, 347 (2010)
  [arXiv:1012.2887 [hep-ph]].



\bibitem{Brodsky:1981kj}
  S.~J.~Brodsky and G.~P.~Lepage,
  Phys.\ Rev.\  D {\bf 24}, 2848 (1981).


\bibitem{Chernyak:1981zz}
  V.~L.~Chernyak and A.~R.~Zhitnitsky,
  Nucl.\ Phys.\  B {\bf 201}, 492 (1982)
  [Erratum-ibid.\  B {\bf 214}, 547 (1983)].


\bibitem{Chernyak:1983ej}
  V.~L.~Chernyak and A.~R.~Zhitnitsky,
  Phys.\ Rept.\  {\bf 112}, 173 (1984).

\bibitem{Mo:2006cy}
  X.~H.~Mo, C.~Z.~Yuan and P.~Wang,
  High Energy Phys. Nucl. Phys. {\bf 31}, 686 (2007)
  [arXiv:hep-ph/0611214].

\bibitem{Zhao:2008eg}
  Q.~Zhao, G.~Li and C.~H.~Chang,
  Chinese Phys. {\bf C 34}, 299 (2010) [arXiv:0812.4092 [hep-ph]].

%
\bibitem{close-amsler} C. Amsler and F.E. Close,
Phys.\ Lett.\ B {\bf 353}, 385 (1995); Phys. Rev. {\bf D53}, 295
(1996).
%
\bibitem{close-kirk} F.E. Close and A. Kirk,
Phys.\ Lett.\ B {\bf 483}, 345 (2000).
%
\bibitem{close-zhao-f0} F.E. Close and Q. Zhao,
Phys. Rev. D {\bf 71}. 094022 (2005) [arXiv:hep-ph/0504043].





\bibitem{Thomas:2007uy}
  C.~E.~Thomas,
  JHEP {\bf 0710}, 026 (2007)
  [arXiv:0705.1500 [hep-ph]].

\bibitem{Escribano:2007cd}
  R.~Escribano and J.~Nadal,
  JHEP {\bf 0705}, 006 (2007)
  [arXiv:hep-ph/0703187].

\bibitem{Cheng:2008ss}
  H.~Y.~Cheng, H.~N.~Li and K.~F.~Liu,
  Phys.\ Rev.\  D {\bf 79}, 014024 (2009)
  [arXiv:0811.2577 [hep-ph]].

\bibitem{Mathieu:2010ss}
  V.~Mathieu and V.~Vento,
  Phys.\ Lett.\  B {\bf 688}, 314 (2010)
  [arXiv:1003.2119 [hep-ph]].



\bibitem{Seiden:1988rr}
  A.~Seiden, H.~F.~W.~Sadrozinski and H.~E.~Haber,
  Phys.\ Rev.\  D {\bf 38}, 824 (1988).


\bibitem{Collaboration:2010kp}
  M. Ablikim {\it et al.}  [BESIII Collaboration],
  arXiv:1012.1117 [hep-ex].



\bibitem{Rosner:2009bp}
  J.~L.~Rosner,
  Phys.\ Rev.\  D {\bf 79}, 097301 (2009)
  [arXiv:0903.1796 [hep-ph]].






\bibitem{Guo:2010ak}
  F.~K.~Guo, C.~Hanhart, G.~Li, U.~G.~Meissner and Q.~Zhao,
  arXiv:1008.3632 [hep-ph].

\bibitem{Guo:2009wr}
  F.~K.~Guo, C.~Hanhart and U.~G.~Meissner,
  Phys.\ Rev.\ Lett.\  {\bf 103}, 082003 (2009)
  [Erratum-ibid.\  {\bf 104}, 109901 (2010)]
  [arXiv:0907.0521 [hep-ph]].


\bibitem{Liu:2009vv}
  X.~H.~Liu and Q.~Zhao,
  Phys.\ Rev.\  D {\bf 81}, 014017 (2010)
  [arXiv:0912.1508 [hep-ph]].

\bibitem{Liu:2010um}
  X.~H.~Liu and Q.~Zhao,
  arXiv:1004.0496 [hep-ph].


\bibitem{Wang:2010iq}
  Q.~Wang, X.~H.~Liu and Q.~Zhao,
  arXiv:1010.1343 [hep-ph].











\end{thebibliography}
\end{document}